\documentclass{appolb}

\begin{document}     
\pagestyle{plain}
\newcount\eLiNe\eLiNe=\inputlineno\advance\eLiNe by -1
\title{{\footnotesize  \hfill  IFUP-TH/2003/21} \\
REVIVAL OF NON-ABELIAN MONOPOLES AND CONFINEMENT IN QCD
}
\author{Kenichi KONISHI
\address{Dipartimento di Fisica, ``E.Fermi", Universit\`a di Pisa, \\
Via Buonarroti, 2, Ed. C  \\
56127 Pisa, Italy}}
\maketitle

\begin{abstract}

Central role played by certain non-Abelian monopoles (of Goddard-Nuyts-Olive-Weinberg type)  
in the infrared dynamics in many confining vacua of softly broken ${\cal N}=2$  supersymmetric gauge theories,  has recently 
been   clarified.  We discuss here the main lessons to be learned from these studies for the confinement nechanism in QCD.

\end{abstract}

\section{Introduction}

Non-Abelian monopoles in spontaneously broken gauge theories have remained a rather obscure 
object for some time now. Apart from the often discussed applications in conformally invariant ${\cal N}=4$ theories
few field theory models were known where such objects play an important role.    A class of ${\cal N}=1$ theories exhibit
 well-known Seiberg's duality; the origin of the ``dual quarks" however remains somewhat mysterious. 

Recent series of work on softly broken ${\cal N}=2 $  gauge theories based on gauge groups $SU(n_c)$, $USp(2n_c)$ and $SO(n_c)$
and with various numbers of flavors, has changed the situation considerably \cite{CKMP}.  It turns out that certain ``dual quarks"
appearing as the low-energy effective degrees of freedom and carrying various non-Abelian charges,  have  the right  properties of the
``semiclassical" non-Abelian monopoles studied earlier, most notably by    Goddard, Nuyts, Olive and  by E. Weinberg \cite{GNO}. 

For example, in  softly broken ${\cal N}=2 $  $SU(n_c)$ theories with $n_f$ flavors,  confining vacua are labelled by an integer $r$, $r=0,1,\ldots,
[{n_f \over 2}]$, which have low-energy effective $SU(r) \times  U(1)^{n_c-r} $ gauge theory description.  The infrared degrees of freedom contain
``dual quarks" carrying charges in the  fundamental representation of the effective $SU(r) $ gauge group, as well as in the fundamental
representation of the flavor
$SU(n_f)$  group.  They carry also a common Abelian charge with respect to one of the $U(1)$ factors.

These are precisely the properties expected for the Goddard-Nuyts-Olive-Weinberg monopoles,  becoming light due to quantum effects,
as  has been shown recently \cite{BK}.  One crucial lesson is that   quantum behavior of  non-Abelian monopoles
depends on  the  massless
flavors   in the original theory,   in an essential manner.

\section{Confinement as non-Abelian dual superconductor}

The importance of the above observation  lies in the fact that in {\it most } of the  ${\cal N}=1$ vacua, confinement is caused by the
condensation of  these non-Abelian monopoles.  Exceptionally  ($r=0$ or $r=1$ vacua of $SU(n_c)$ theory)    the low-energy theory is an
Abelian magnetic gauge theory and  confinement is described as a dual Meissner effect, as proposed by 't Hooft for QCD \cite{TH}. 
However,  confinement  in generic $r$-vacua  is a dual superconductivity of non-Abelian variety.

The fact that such $r$-vacua appear only for $r < {n_f \over 2}$  can be understood as an effect of renormalization:  only for these 
vacues of $r$,  the low-energy $SU(r)$ gauge group is infrared free, with the monopoles carrying flavor charges of the fundamental quarks. 
The beta function of the dual, magnetic theory has an opposite sign with respect to that of the electric $SU(n_c)$ theory,
\begin{equation}   b_0^{(dual)} \propto   -  2 \, r  +   n_f  >  0,   
\qquad       b_{0} \propto  -  2 \, n_c +    n_f  <     0,      
\label{betafund}  \end{equation} 
and this reflects a particular property of ${\cal N}=2$ gauge theory  with a small coefficient ($2$)  in front of the color multiplicity in 
$b_0$.

\section{Deformed conformal vacua and confinement }   

For this reason, it is not surprising that the most typical set of vaua in confinement phase in the class of models studied in 
\cite{CKMP}  turn out to be based, rather, on a nontrivial superconformal theory \footnote{In contrast, the generic $r$-vacua are trivial -
infrared free - superconformal theories.}.  Examples are the
$r={n_f \over 2}$   vacua of
$SU(n_c)$ theory and {\it all} of confining  vacua of $USp(2n_c)$  and $SO(n_c)$ theories with vanishing bare quark masses. 
${\cal N}=1$ perturbation -  nonzero adjoint matter mass which triggers dual Higgs mechanism -  gives  a deformation 
of such infrared fixed-point theories. 
Low-energy effective theory contains relatively non-local set of gauge and matter fields carrying non-Abelian charges, 
and no simple local field theory description is available.  This makes the analysis of these vacua a difficult task. 
 A first step to study
these cases more closely was undertaken in
\cite{AGK}, by considering a concrete example of
$r=2$ vacua  of softly broken $SU(3)$ gauge theory with four quark flavors.   This study indicates that  the confinement 
is a dual (non-Abelian) superconductor,  but that the condensation of the monopoles is a strong interaction phenomenon, rather than 
a (dual) perturbative mechanism as in the $r < {n_f \over 2}$   vacua.

\section{QCD} 

What can one learn from these studies in supersymmetric theories about the confinement mechanism in the real-world  QCD?
Here  we know 

\noindent  {(i)}  that no dynamical Abelianization  occurs; 

\noindent  {(ii)}    that,   on the other hand,   in QCD  with $n_f$  flavor,     the original and dual beta functions have the first coefficients
($n_c=3$, ${\tilde n}_c=2,3$)
\begin{equation}    b_0=   - 11 \, n_c +  2 \,n_f \quad  {\hbox {\rm vs}} \quad   {\tilde b}_0  =  - 11 \, {\tilde n}_c +   n_f:   
\end{equation}  they have the same sign  because of the large coefficient  in front of the color  multiplicity  ({\it cfr.}
Eq.(\ref{betafund})).

Barring that higher loops change the situation, this leaves us with  the option of   strongly-interacting non-Abelian monopoles,
somewhat like in the cases discussed in 3.  Is it possible that non-Abelian monopoles (perhaps certain  composite
theirof)  carrying nontrivial flavor
$SU_L(n_f)\times SU_R(n_f)$  quantum numbers condense 
yielding  the global   symmetry breaking such as 
$   G_F = SU_L(n_f)\times SU_R(n_f) \Rightarrow   SU_V(3),   $
observed in Nature? 
How are   't Hooft's Abelian monopoles   related to   these non-Abelian  monopoles?   These are the questions to be studied further. 

A  more detailed account of  these discussions  appeared in \cite{Konishi}.


\begin{thebibliography}{100}

\bibitem{CKMP}  G. Carlino, K. Konishi and H. Murayama,
 {\bf    Nucl. Phys.  B590}  (2000) 37,     hep-th/0005076; 
G. Carlino, K. Konishi, P. S.  Kumar  and H. Murayama,
 hep-th/0104064,   {\bf    Nucl. Phys.  B608}  (2001) 51.


\bibitem{GNO}   P. Goddard, J. Nuyts and D. Olive,   {\bf Nucl. Phys.  B125}
(1977) 1,    E. Weinberg, {\bf Nucl. Phys. B167} (1980) 500;  {\bf Nucl. Phys. B203} (1982) 445; 
``Massive and Massless Monopoles and Duality",  hep-th/9908095. 
  

\bibitem{BK}   S. Bolognesi and K. Konishi,  {\bf    Nucl. Phys.  B645 }  (2002) 337, hep-th/0207161. 


\bibitem{TH}  G. 't Hooft, {\bf  Nucl. Phys.   B190}   (1981) 455;
S. Mandelstam,  {\bf Phys. Lett.  53B }  (1975) 476.  


\bibitem{AGK}  R. Auzzi, R. Grena and K. Konishi,  {\bf    Nucl. Phys.  B653 } (2003) 204, hep-th/0211282.

\bibitem{Konishi}  K. Konishi,  ``Who Confines Quarks? - On Non-Abelian Monopoles and Dynamics of Confinement",  hep-th/0304157. 

          

\end{thebibliography}
\end{document}